\documentclass[12pt]{article}

\usepackage[a4paper,margin=2.5cm]{geometry}
\usepackage{graphicx}%
\usepackage{amsmath,amssymb,amsfonts}%
\usepackage{amsthm}%
\usepackage{hyperref}
\usepackage{booktabs}%
\usepackage{cite}
\usepackage{array}
\usepackage{setspace}
\begin{document}
% \linenumbers
% \modulolinenumbers[1]
%TC:ignore
\title{\textbf{Symmetry-guided and AI-accelerated design of intercalated transition metal dichalcogenides for antiferromagnetic spintronics}}
\author{
\small
\parbox{\textwidth}{
\centering
Yu Pang$^{1,\dagger}$, Yue Gu$^{1,\dagger}$, Runsheng Zhong$^{1,\dagger}$, Liyang Zou$^{1,\dagger}$, Xiaobin Chen$^{2}$, Xiaolong Zou$^{1,*}$, Wenhui Duan$^{3,4,5}$ \\
$^{1}$Shenzhen Geim Graphene Center \& Shenzhen Key Laboratory of Advanced Layered Materials for Value-added Applications, Institute of Materials Research, Tsinghua Shenzhen International Graduate School, Tsinghua University, Shenzhen 518055, China\\
$^{2}$School of Science, State Key Laboratory on Tunable Laser Technology and Ministry of Industry and Information Technology Key Lab of Micro-Nano Optoelectronic Information System, Harbin Institute of Technology, Shenzhen, Shenzhen 518055, China\\
$^{3}$State Key Laboratory of Low Dimensional Quantum Physics and Department of Physics, Tsinghua University, Beijing 100084, China\\
$^{4}$Institute for Advanced Study, Tsinghua University, Beijing 100084, China\\
$^{5}$Frontier Science Center for Quantum Information, Beijing 100084, China\\
$^{\dagger}$These authors contributed equally to this work.\\
$^{*}$Corresponding author: xlzou@sz.tsinghua.edu.cn
}
}
\date{}
\maketitle
% \vspace{-0.5cm}

\noindent\textbf{Abstract:}
The advancement of antiferromagnetic spintronics depends on quantum materials with target symmetry-dictated functionalities, however, their systematic discovery is hindered by the immense configurational complexity of the available material space. Here, we introduce a symmetry-guided, AI-accelerated framework incorporating graph neural networks with high generalization ability to overcome this bottleneck. Based on fully intercalated transition metal dichalcogenides (iTMDs) and using only 200 relaxed partially intercalated structures for transfer learning, our model effectively explores more than 100,000 partially intercalated configurations and identifies 35 altermagnetic and 20 $\mathcal{T}\tau$-antiferromagnetic ground-state candidates. Interestingly, we show that tuning spin-group symmetry through intercalant arrangement or magnetic ordering realizes a series of d-wave altermagnets in these hexagonal systems with high spin-charge conversion efficiency. Furthermore, we reveal plentiful $\mathcal{T}\tau$-antiferromagnets enabling efficient N\'eel spin-orbit torque switching, driven by giant $\mathcal{T}$-odd spin Edelstein susceptibilities. These results establish iTMDs as a versatile platform for spintronics and provide a general strategy for the accelerated design of symmetry-enforced quantum materials.
% \keywords{Intercalated TMDs, Altermagnet, Spin-Orbit Torque, Symmetry analysis, Multi-task graph neural network}

\maketitle
%TC:endignore
%%=============================================================%%
%%  正文 (Main Text)
%%=============================================================%%

\section{Introduction}\label{sec1}
The development of high-density, high-speed spintronic devices is fundamentally hindered by the intrinsic limitations of their two foundational material classes. Ferromagnets (FMs) are easily controlled but suffer from stray fields, while conventional antiferromagnets (AFMs) offer ultrafast dynamics and zero stray fields but are notoriously difficult to  manipulate~\cite{jungwirthAntiferromagneticSpintronics2016,baltzAntiferromagneticSpintronics2018,hanCoherentAntiferromagneticSpintronics2023}. Two main strategies have emerged to resolve this challenge. The first focuses on enabling electrical control of conventional AFMs, primarily through N\'eel spin-orbit torque (SOT). $\mathcal{T}\tau$ AFMs lacking local inversion symmetry, which can generate efficient anti-damping N\'eel torques, are representative materials to achieve such a goal~\cite{zeleznyRelativisticNeelOrderFields2014,zeleznySpinorbitTorquesLocally2017,manchonCurrentinducedSpinorbitTorques2019}. The second paradigm seeks a more fundamental solution by introducing an unconventional AFM class that inherently combines the advantages of both conventional FMs and AFMs, called altermagnets (AMs)~\cite{smejkalConventionalFerromagnetismAntiferromagnetism2022,smejkalEmergingResearchLandscape2022,yuanGiantMomentumdependentSpin2020,wuFermiLiquidInstabilities2007,maMultifunctionalAntiferromagneticMaterials2021a,shaoSpinneutralCurrentsSpintronics2021b,baiAltermagnetismExploringNew2024,krempaskyAltermagneticLiftingKramers2024,xuChemicalDesignMonolayer2026,gaoAIacceleratedDiscoveryAltermagnetic2025,cheEngineeringAltermagneticStates2025,liuRobustAltermagnetismCompensated2025,guoTunableAltermagnetismInterchain2025,heNonrelativisticSpinMomentumCoupling2023,liuTwistedMagneticVan2024,zengDescriptionTwodimensionalAltermagnetism2024,panGeneralStackingTheory2024a,jeongAltermagneticPolarMetallic2026}. AMs are a distinct magnetic class characterized by a compensated net magnetization and a non-relativistic, momentum-dependent spin splitting in their electronic bands. This unique property originates from crystal symmetries that connect the two opposite spin sublattices via a rotational or mirror operation, enabling robust electrical detection and control of magnetism without the penalty of stray fields. Furthermore, these symmetry properties give rise to exotic spintronic phenomena, including the spin-splitter effect\cite{baiEfficientSpintoChargeConversion2023,gonzalez-hernandezEfficientElectricalSpin2021,laiWaveFlatFermi2025}, the anomalous Hall effect\cite{smejkalCrystalTimereversalSymmetry2020}, piezomagnetism\cite{maMultifunctionalAntiferromagneticMaterials2021a}, magneto-optic responses\cite{panExperimentalEvidenceNeelOrderDriven2026,liuUncompensatedLinearDichroism2025,liUnconventionalMagnetoOpticalEffects2025} and non-relativistic altermagnetic multiferroicity\cite{huangSpinInversionEnforced2025,dingFerroelasticallyTunableAltermagnets2025,duanAntiferroelectricAltermagnetsAntiferroelectricity2025,guFerroelectricSwitchableAltermagnetism2025,sunUnifiedSymmetryFramework2026,liFerrovalleyPhysicsStacked2025} just to name a few. These features position AMs as promising candidates for next-generation spintronic applications. Great efforts have been devoted to search for AM candidates, through group theory analysis\cite{smejkalConventionalFerromagnetismAntiferromagnetism2022,chenEnumerationRepresentationTheory2024,xiaoSpinSpaceGroups2024,jiangEnumerationSpinSpaceGroups2024}, chemical design of known AMs\cite{xuChemicalDesignMonolayer2026}, and  machine learning classifier\cite{gaoAIacceleratedDiscoveryAltermagnetic2025}. However, both altermagnets and antiferromagnets supporting N\'eel SOT switching are subject to stringent symmetry constraints, which restrict potential candidates to a narrow subset of the vast configuration space. Consequently, experimentally realized examples remain scarce, underscoring the urgent need for symmetry-guided frameworks to accelerate material discovery.

Intercalated transition metal dichalcogenides (iTMDs) with hexagonal lattices host diverse emergent quantum phases, including charge density waves, unconventional superconductivity, and correlated insulating states\cite{liIntercalationdrivenTunabilityTwodimensional2024}. Recent experiments have synthesized diverse iTMDs with intercalant concentration up to 100\% through different growth methods\cite{zhaoEngineeringCovalentlyBonded2020,zhou2025cation}. By tuning the chemical composition, structural order, and magnetic interaction, they serve as a fertile material platform for antiferromagnetic spintronics, such as antiferromagnetic $\mathrm{Fe{_{1/3}}NbS_2}$ for electrically switching the N\'eel vector and the non-coplanar antiferromagnets $\mathrm{Co_{1/3}Ta_3S_6}$ and $\mathrm{Co_{1/3}Nb_3S_6}$ exhibiting a spontaneous topological Hall effect\cite{takagiSpontaneousTopologicalHall2023}. 
% , \textit{i.e.}, AMs in particular those going beyond the currently observed g-wave ones with C$_3$ symmetry~\cite{spragueObservationAltermagneticSpin2025,daleNonrelativisticSpinSplitting2024,grahamLocalProbeEvidence2025,sakhyaElectronicStructureLayered2025,regmiAltermagnetismLayeredIntercalated2025} and $\mathcal{T}\tau$ AFMs that enable intrinsic N\'eel SOT~\cite{wadleyElectricalSwitchingAntiferromagnet2016,grzybowskiImagingCurrentInducedSwitching2017,godinhoElectricallyInducedDetected2018}. 
Realizing AMs and $\mathcal{T}\tau$ AFMs within this universal platform could facilitate the deep understanding of unconventional magnetism and its interplay with diverse quantum phases.
For altermagnetism, since hexagonal symmetry is prevalent among experimentally realized layered materials~\cite{manzeli2DTransitionMetal2017,chhowallaChemistryTwodimensionalLayered2013}, the only two AMs experimentally discovered in this platform (e.g., $\mathrm{Co{_{1/4}}TaSe_2}$~\cite{spragueObservationAltermagneticSpin2025} and $\mathrm{Co{_{1/4}}NbSe_2}$~\cite{daleNonrelativisticSpinSplitting2024,grahamLocalProbeEvidence2025,sakhyaElectronicStructureLayered2025,regmiAltermagnetismLayeredIntercalated2025}) are g-wave, limited to structures preserving C$_3$ symmetry. 
 Recent work on $\mathrm{Co_xNbSe_2}$ further revealed that local intercalant ordering and synthesis history can drive an evolution from the AM regime near $x=1/4$ to spin-glass-like and spin-density-wave phases, underscoring the rich and highly tunable magnetic landscape of iTMDs~\cite{mandujanoEvolutionAltermagnetismSpin2025}.
However, a multitude of $C_3$-breaking configurations, arising from interlayer stacking, intercalant arrangement or magnetic ordering, should be physically accessible at fractional concentrations. These configurations would readily stabilize the functionally pivotal d-wave phases, yet they remain largely unexplored. Furthermore, for N\'eel-vector-switching AFMs, the presence of heavy elements in the TMD family provides the strong spin-orbit coupling essential for efficient N\'eel SOT~\cite{nairElectricalSwitchingMagnetically2020}. 
% Despite this clear potential and the well-established switching in archetypal AFMs like CuMnAs~\cite{wadleyElectricalSwitchingAntiferromagnet2016,grzybowskiImagingCurrentInducedSwitching2017,godinhoElectricallyInducedDetected2018}, its realization in the intercalated TMD family remains ambiguous, with the few examples, such as $\mathrm{Fe{_{1/3}}NbS_2}$, complicated by coexisting complex magnetic orders~\cite{nairElectricalSwitchingMagnetically2020,manivAntiferromagneticSwitchingDriven2021}. 
The theoretical identification of both AMs and N\'eel-vector-switching AFMs in iTMDs hinges on finding stable configurations that satisfy specific symmetry requirements. The purposeful design of target materials necessitates a systematic exploration of configuration space with vast chemical, structural, and magnetic order tunability, which poses significant challenges. Accordingly, a paradigm shift towards systematic and data-driven discovery methods is called for.  

Here, we establish a general symmetry-guided computational framework combining group theory, high-throughput density functional theory (DFT) calculations, and an in-house developed multi-task graph neural network (MT-GNN) with improved generalization ability for exploring vast configuration space, and apply this framework to the notoriously complex landscape of intercalated hexagonal layered TMDs for searching promising antiferromagnetic spintronic candidates. Our screening in the fully intercalated systems uncovers a comprehensive repository of 65 altermagnets and 94 candidates for electrically switchable $\mathcal{T}\tau$-AFMs. By employing transfer learning with only 200 randomly selected partially iTMDs, we explore the even more complex configuration space of partial intercalation with over 100,000 structures, and identify additional 154 AMs and 149 $\mathcal{T}\tau$-AFMs, which includes 35 AM and 20 $\mathcal{T}\tau$-AFM ground-state structures with high experimental synthesizability, complemented by a broad library of 119 AM and 129 $\mathcal{T}\tau$-AFM metastable candidates. Importantly, the intercalation provides a feasible route to control spin group symmetry through interlayer stacking, intercalant arrangement, and magnetic ordering, leading to a series of d-wave altermagnets in this hexagonal platform. Simultaneously, $\mathcal{T}\tau$-AFMs supporting $\mathcal{T}$-odd Néel SOT originating from an intrinsic contribution driven by impurity-independent interband transitions have been revealed in both metallic and insulating systems, overcoming the strict conductivity constraints of $\mathcal{P}\mathcal{T}$-symmetric systems. 
%By further integrating a machine-learning-accelerated approach to navigate the partially intercalated regime, we identify an additional 35 AMs and 20 $\mathcal{T}\tau$-AFMs, crucially uncovering d-wave candidates that demonstrate the accessibility of this symmetry in the hexagonal platform.

\section{Results}\label{sec2}
\subsection{A systematic search workflow for target magnetic phases}
\begin{figure}[htb]
	\centering
	\includegraphics[width=\linewidth]{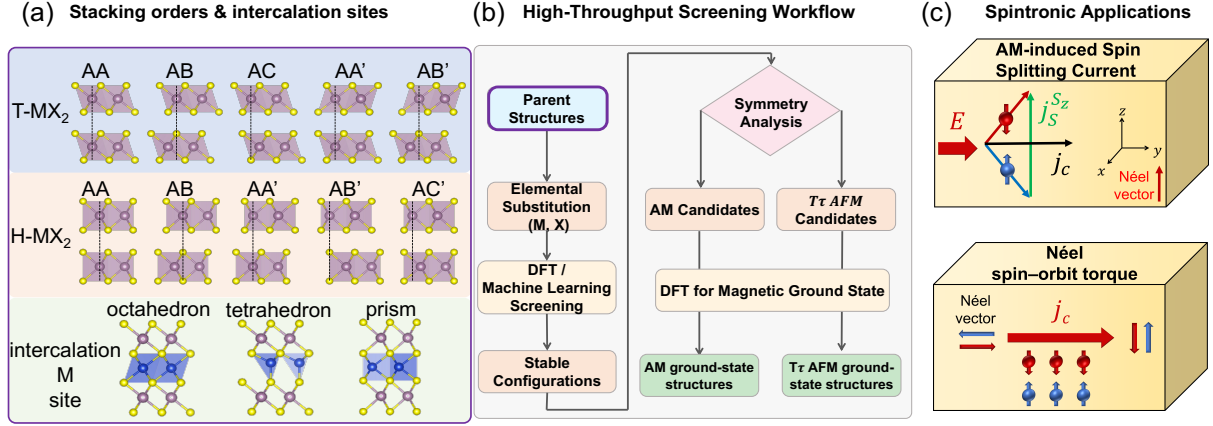}
\caption{\textbf{A symmetry-guided and AI-accelerated workflow for discovering target magnetic phases in iTMDs.}
\textbf{a}, Schematic configurations in both T- and H-phases considering different stacking orders and intercalation positions including octahedral, tetrahedral, and trigonal prismatic sites. \textbf{b}, The screening workflow adopted to identify stable altermagnetic and $\mathcal{T}\tau$-antiferromagnetic ground states, by combining first-principles calculations, machine learning, and symmetry analysis.  \textbf{c}, The target spintronic functionalities arising from distinct symmetries, including the AM-induced spin-splitting current and the N\'eel-vector-switching SOT in $\mathcal{T}\tau$-AFMs.}
	\label{fig:1}
\end{figure}

Our search for target magnetic phases begins by systematically exploring fully iTMDs. We select two MX$_2$ monolayer polymorphs (T- and H-phases) and consider five stacking orders for each, as depicted in Fig.~\ref{fig:1}a. For each of the resulting ten configurations, we then identify three high-symmetry sites for intercalation, \textit{i.e.}, octahedral, tetrahedral, and trigonal prismatic. The combination of these features yields 26 distinct geometric prototypes. We further note that for three prototypes involving tetrahedral intercalation (T-AB'-tetra, H-AA'-tetra, and H-AB'-tetra), two symmetrically inequivalent configurations exist depending on the relative alignment of two intercalated layers. Accounting for these subtleties results in an exhaustive library of 29 unique structural prototypes, covering both pristine hosts and fully intercalated phases. (Table~\ref{tab:full_intercalation_prototypes} and Supplementary Note S1.1). %This initial library not only defines the search space for our work on full intercalation, but also provides the essential building blocks for tackling even more complex landscape of partial intercalation.

To overcome combinatorial complexity, we design a symmetry-guided, multi-stage high-throughput screening workflow illustrated in Fig.~\ref{fig:1}b. First, for 29 parent prototypes, we perform extensive elemental substitution, sampling a wide range of transition metals (spanning the 3$d$, 4$d$, and 5$d$ series) and chalcogens (S, Se, Te), as detailed in Supplementary Note S1.2. The thermodynamic ground state for each chemical composition is then identified using an integrated computational strategy. While direct DFT calculations are employed for the fully intercalated systems, a MT-GNN accelerated approach, adapted from our prior work~\cite{liang2024multi}, is employed to efficiently explore the much larger and computationally demanding phase space of partially intercalated structures. Our MT-GNN based on DimeNet++ and mixture density networks show enhanced performance in predicting materials properties from unrelaxed configurations and improved generalization ability to unseen out-of-domain samples (see Methods and details in Section~\ref{sec:ML}).

Second, after stable structural candidates identified, we leverage group theory to search over all possible magnetic orderings. Altermagnetic candidates are identified based on their spin group symmetries, requiring a rotational operation to connect the two opposite spin sublattices in the absence of $P\mathcal{T}$ or $\mathcal{T}\tau$ symmetry. For N\'eel-vector-switching candidates, we focus on $\mathcal{T}\tau$-symmetric AFMs with broken spatial inversion symmetry, which allows for a $\mathcal{T}$-odd spin Edelstein effect, generating highly efficient anti-damping SOT.
 % We reinforce inversion symmetry breaking in our group theory analysis to pre-select candidates with potential to host the desired physics, instead of a brute-force screening. 
%While the essential ingredient for this effect is the breaking of local inversion symmetry at the magnetic sites, we implement a stricter and more computationally tractable criterion, \textit{i.e.}, 

Finally, for all promising candidates, we perform DFT calculations to determine their magnetic ground state. This stage evaluates total energies for a set of collinear magnetic configurations, encompassing various intralayer (FM, stripe-AFM, zigzag-AFM) and interlayer magnetic orderings (Supplementary Note S2). Our multi-stage process yields a repository of functional materials, identifying 65 AMs and 94 $\mathcal{T}\tau$-AFMs in the fully intercalated systems, and additional 154 AMs and 149 $\mathcal{T}\tau$-AFMs at 1/2 and 1/4 concentrations, including 35 AMs and 20 $\mathcal{T}\tau$-AFMs ground-state candidates for further experimental verification (Supplementary Note S9). 

This repository of materials shows good potential for target spintronics applications. 
Among the identified altermagnets, d-wave candidates are particularly compelling as their symmetry enables the spin-splitter effect\cite{gonzalez-hernandezEfficientElectricalSpin2021} shown in the top panel of Fig.~\ref{fig:1}c. This mechanism generates a transverse spin current with spin polarization collinear with the N\'eel vector, providing an all-electrical route for the magnetic-field-free switching of adjacent magnetic layers with perpendicular magnetic anisotropy (PMA). For the discovered $\mathcal{T}\tau$-AFMs, current-induced spin accumulation via the spin Edelstein effect exerts an anti-damping torque on the magnetic sublattices that deterministically switches the N\'eel vector, displayed in the bottom panel of Fig.~\ref{fig:1}c.
% Among the altermagnets identified, d-wave candidates are particularly compelling, as their specific symmetry enables the spin-splitter effect. As shown in the upper panel of Fig.~\ref{fig:1}c, an applied electric field ($E$) drives a longitudinal charge current ($j_c$), which in turn generates a pure transverse spin current ($j_s$). Crucially, the out-of-plane spin polarization ($S_z$) of this current, which is collinear to the N\'eel vector, provides an all-electrical route for achieving the field-free switching of adjacent magnetic layers with perpendicular magnetic anisotropy (PMA), a key goal for next-generation memory technologies. The discovered $\mathcal{T}\tau$-AFMs, on the other hand, are primed for efficient electrical switching of the N\'eel vector itself. In these materials, a charge current induces a non-equilibrium spin polarization via the spin Edelstein effect, which exerts an anti-damping SOT on the magnetic sublattices, enabling deterministic control over the magnetic order. 
In the following sections, we first present in-depth case studies of one representative fully intercalated candidate from each class, followed by the extension of our AI-accelerated framework to the complex configuration space of partial intercalation.

\subsection{d-wave altermagnets for significant spin-splitter current}
\begin{figure}[htb]
	\centering
	\includegraphics[width=\linewidth]{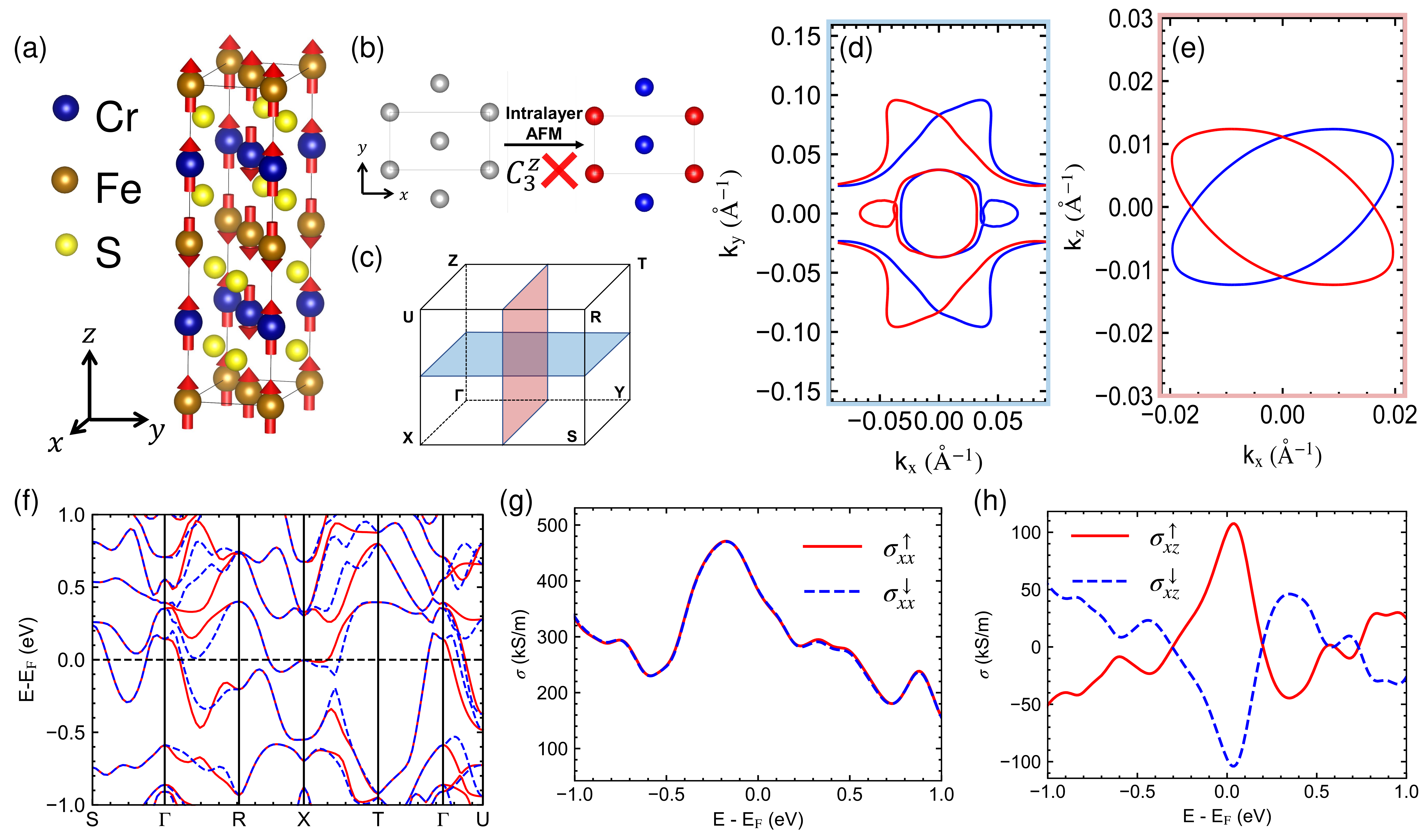}
	\caption{\textbf{A d-wave altermagnet with significant spin-splitter conductivity.} 
	\textbf{a}, Crystal and magnetic structure of the d-wave altermagnet candidate H-AA'-octa Fe-CrS$_2$.
	\textbf{b}, Schematic showing the breaking of the $C_3$ rotational symmetry by intralayer antiferromagnetic order of the Cr atoms .
	\textbf{c}, The first Brillouin zone. The blue plane at constant $k_z$ and the pink plane at constant $k_y$ correspond to the momentum-space cuts for the Fermi surfaces shown in panels (\textbf{d}) and (\textbf{e}), respectively.
	\textbf{d}, \textbf{e}, Spin-resolved 2D Fermi surfaces at two different momentum-space planes, showcasing strong anisotropy and the two-fold symmetry characteristic of a d-wave state. Red and blue lines represent spin-up and spin-down channels.
	\textbf{f}, Non-relativistic band structure along high-symmetry paths, highlighting the large, momentum-dependent spin splitting. Solid red and dashed blue lines denote spin-up and spin-down bands.
	\textbf{g}, \textbf{h}, Calculated longitudinal ($\sigma^\uparrow_{xx}$, $\sigma^\downarrow_{xx}$) and transverse ($\sigma^\uparrow_{xz}$, $\sigma^\downarrow_{xz}$) spin conductivities as a function of Fermi energy, respectively.}
	\label{fig:AMdemo}
\end{figure}

Fully intercalated Fe-CrS$_2$ in the hexagonal H-AA'-octa structure is a representative d-wave altermagnet. While Fe atom layers feature intralayer FM and interlayer AFM coupling, the two Cr atom layers adopt intralayer stripe AFM and interlayer FM coupling (Fig.~\ref{fig:AMdemo}a). The stripe AFM ordering of Cr atoms breaks $C_3$ rotational symmetry, which is critical for d-wave altermagnetism (Fig.~\ref{fig:AMdemo}b). 
Symmetry analysis reveals that the magnetic structure is described by the spin point group $^2m^2m^1m$, comprising eight symmetry elements. Four operations, $\{E||E\}$, $\{E||P\}$, $\{E||M_y\}$, and $\{E||C_{2y}\}$, transform magnetic sublattices into themselves, while the other four, $\{C_2||M_x\}$, $\{C_2||C_{2x}\}$, $\{C_2||M_z\}$, and $\{C_2||C_{2z}\}$,  combined with a fractional lattice translation $\tau = (0.5, 0.5, 0.5)$, connect the spin-up and spin-down sublattices, where $C_2$ is a 180-degree spin rotation. These altermagnetic symmetries impose strict constraints on the electronic band structure. Specifically, the $\{C_2||M_x\}$ symmetry dictates that $\epsilon^\uparrow(n, k_x, k_y, k_z) = \epsilon^\downarrow(n, -k_x, k_y, k_z)$, enforcing spin degeneracy on the $k_y-k_z$ plane (\textit{i.e.}, $k_x=0$). Similarly, the $\{C_2||M_z\}$ symmetry leads to $\epsilon^\uparrow(n, k_x, k_y, k_z) = \epsilon^\downarrow(n, k_x, k_y, -k_z)$, resulting in spin degeneracy across the $k_x-k_y$ plane (\textit{i.e.}, $k_z=0$).

% Symmetry analysis reveals that the magnetic structure is described by a spin group $\{U||R|t\}$, where $U$ is a spin rotation, $R$ is a point group operation, and $t$ is a fractional translation. Four operations, $\{E||E\}$, $\{E||P\}$, $\{E||M_y\}$, and $\{E||C_{2y}\}$, transform magnetic sublattices into themselves, while the other four, $\{C_2||M_x|\tau\}$, $\{C_2||C_{2x}|\tau\}$, $\{C_2||M_z|\tau\}$, and $\{C_2||C_{2z}|\tau\}$, connect the spin-up and spin-down sublattices, where $C_2$ is a 180-degree spin rotation and $\tau = (0.5, 0.5, 0.5)$ is a fractional lattice translation. These altermagnetic symmetries impose strict constraints on the electronic band structure. Specifically, the $\{C_2||M_x|\tau\}$ symmetry dictates that $\epsilon^\uparrow(n, k_x, k_y, k_z) = \epsilon^\downarrow(n, -k_x, k_y, k_z)$, enforcing spin degeneracy on the $k_y-k_z$ plane (\textit{i.e.}, $k_x=0$). Similarly, the $\{C_2||M_z|\tau\}$ symmetry leads to $\epsilon^\uparrow(n, k_x, k_y, k_z) = \epsilon^\downarrow(n, k_x, k_y, -k_z)$, resulting in spin degeneracy across the $k_x-k_y$ plane (\textit{i.e.}, $k_z=0$).

Fig.~\ref{fig:AMdemo}f displays the calculated band structure without spin-orbit coupling (SOC), revealing a $k$-dependent spin splitting that is the defining characteristic of altermagnets. In agreement with our symmetry analysis, the bands remain spin-degenerate along the paths S-$\Gamma$, R-X, and T-$\Gamma$. The d-wave nature of the spin splitting is visualized in the spin-resolved iso-energy surfaces, shown for two different momentum-space planes within the first Brillouin zone (Figs.~\ref{fig:AMdemo}d,e). These surfaces exhibit strong anisotropy and a two-fold rotational symmetry, different from the six-fold or three-fold symmetry expected in hexagonal systems. This confirms that the breaking of $C_3$ symmetry by the magnetic order gives rise to a robust d-wave altermagnetic state.

To quantify the spin-splitter effect, we calculate the spin conductivity tensor, $\sigma^{s}$, using the Boltzmann transport equation under the constant relaxation time approximation.
Given the net spin conductivity is defined as $\sigma = \sigma^{\uparrow} - \sigma^{\downarrow}$, the only non-zero components of the spin conductivity tensor in this system are $\sigma_{xz}$ and its symmetric counterpart $\sigma_{zx}$. Accordingly, a charge current applied along the $x$-direction generates a pure spin current flowing in the $z$-direction, with spins polarized collinearly with the N\'eel vector. This mechanism is ideal for achieving field-free switching of materials with PMA.

The calculated longitudinal ($\sigma_{xx}$) and transverse ($\sigma_{xz}$) components of the spin conductivity are plotted as a function of Fermi energy in Fig.~\ref{fig:AMdemo}g and \ref{fig:AMdemo}h, respectively, which perfectly match our symmetry predictions. Notably, the transverse component of the net spin conductivity, $\sigma_{xz}$, exhibits a strong peak of approximately 100~kS/m at the Fermi level, an order of magnitude larger than that of Mn$_5$Si$_3$ ($\sim$10~kS/m\cite{dongFieldfreePerpendicularMagnetization2025}). We further quantify the efficiency by the spin-splitting angle, $\alpha_{zx} = |\frac{\sigma_{zx}^{\uparrow}-\sigma_{zx}^{\downarrow}}{\sigma_{xx}^{\uparrow}+\sigma_{xx}^{\downarrow}}|$, which reaches a substantial value of approximately 26\%, comparable to that of Mn$_5$Si$_3$\cite{dongFieldfreePerpendicularMagnetization2025}. %This combination of a significant spin-splitter conductivity and a large efficiency establishes Fe-CrS$_2$ as a premier candidate for generating and manipulating collinear spin currents.

\subsection{Electrically switchable \texorpdfstring{$\mathcal{T}\tau$}-AFMs}
\begin{figure}[htb]
	\centering
	\includegraphics[width=\linewidth]{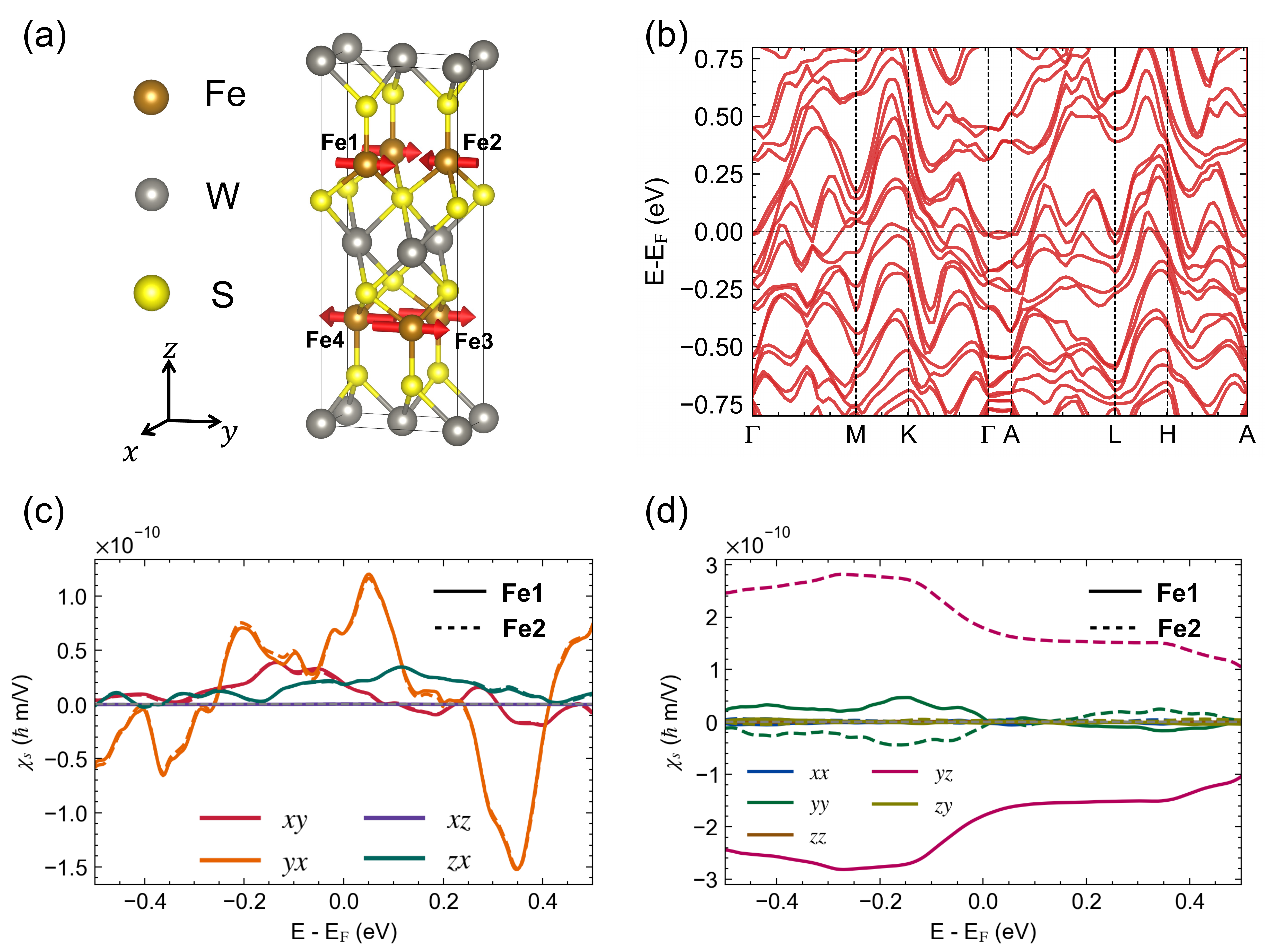}
    \caption{\textbf{An electrically switchable $\mathcal{T}\tau$-AFMs with a strong spin Edelstein effect.} 
    \textbf{a}, Crystal and magnetic structures of the $\mathcal{T}\tau$-symmetric AFM candidate, T-AB'-tetra\_2 Fe-WS$_2$. The two symmetry-related magnetic Fe atoms in the unit cell are labeled as Fe1 and Fe2.
    \textbf{b}, Relativistic band structure calculated with SOC, revealing a metallic ground state.
    \textbf{c, d}, Calculated components of the local spin Edelstein effect susceptibility tensor projected onto the Fe1 (solid lines) and Fe2 (dashed lines) atoms as a function of Fermi energy. 
    (\textbf{c}) The $\mathcal{T}$-even components ($\chi_{xy}, \chi_{yx}, \chi_{xz}, \chi_{zx}$) exhibit identical responses for both sublattices. 
    (\textbf{d}) The $\mathcal{T}$-odd components ($\chi_{xx}, \chi_{yy}, \chi_{zz}, \chi_{yz}, \chi_{zy}$) exhibit responses of equal magnitude but opposite sign, consistent with the $\mathcal{T}\tau$ symmetry constraints and confirming the origin of the staggered N\'eel torque.}
	\label{SOTdemo}
\end{figure}

%--------- Section 3a-b: Material and Symmetry ---------
As an example of electrically switchable $\mathcal{T}\tau$-AFMs, we analyze fully intercalated Fe-WS$_2$ in the T-AB'-tetra\_2 structure (Fig.~\ref{SOTdemo}a). Two Fe layers exhibit an intralayer stripe AFM order that breaks the C$_3$ rotational symmetry, resulting in a magnetic space group $P_Ac$. %The two opposite spin sublattices are connected by the combined symmetry operation $\mathcal{T}\tau$, where $\mathcal{T}$ is time reversal and $\tau = (1/2, 1/2, 0)$ is a fractional lattice translation. 
The corresponding magnetic point group, composed of four symmetry operations, is identified as $m.1'$. 
%To assess its potential for electrical switching, we first determine its magnetic anisotropy via DFT calculations including SOC. 
The ground state features a N\'eel vector oriented along the in-plane $y$-direction, with in-plane and out-of-plane magnetic anisotropy energies of 1.0 and 3.4~meV, respectively. The calculated band structure (Fig.~\ref{SOTdemo}b) confirms that Fe-WS$_2$ is metallic, with a high density of states at the Fermi level, suggesting its potential for a strong current-induced spin response.
%--------- Section 3c-d: Spin Edelstein Effect ---------

The key mechanism for electrical control of magnetization in a $\mathcal{T}\tau$-AFM is the spin Edelstein effect (SEE), where an applied electric field $\mathbf{E}$ induces a non-equilibrium spin polarization $\delta\mathbf{S}$, governed by the linear response tensor $\chi_{ij}$ ($\delta S_i = \chi_{ij} E_j$). %This response can be decomposed into $\mathcal{T}$-even ($\chi^{\text{even}}$) and $\mathcal{T}$-odd ($\chi^{\text{odd}}$) components, which correspond to the intraband and interband contributions, respectively (see Methods for the full Kubo formalism). The symmetry of the magnetic crystal imposes strict selection rules on the non-zero elements of the $\chi$ tensor. 
For the $m.1'$ magnetic point group of Fe-WS$_2$, our analysis predicts four non-zero $\mathcal{T}$-even components for each sublattice ($\chi_{xy}$,$\chi_{yx}$,$\chi_{xz}$, and $\chi_{zx}$) and five non-zero $\mathcal{T}$-odd components ($\chi_{xx}$,$\chi_{yy}$,$\chi_{zz}$,$\chi_{yz}$, and $\chi_{zy}$). Furthermore, the $\mathcal{T}\tau$ symmetry relating the two sublattices dictates that their $\mathcal{T}$-even responses must be identical ($\chi^{\uparrow}_{\text{even}} = \chi^{\downarrow}_{\text{even}}$), whereas their $\mathcal{T}$-odd responses must be opposite ($\chi^{\uparrow}_{\text{odd}} = -\chi^{\downarrow}_{\text{odd}}$). Our first-principles calculations of the local SEE tensor components (see Methods for the full Kubo formalism) corroborate the symmetry analysis. As illustrated in Fig.~\ref{SOTdemo}c and \ref{SOTdemo}d, the $\mathcal{T}$-even components (e.g., $\chi_{yx}$) exhibit identical spectral features for both Fe1 and Fe2 atoms, while the $\mathcal{T}$-odd components are equal in magnitude but opposite in sign, with the $\chi_{yz}$ component exhibiting the dominating response. Results for the remaining Fe atoms, which are related by the $\mathcal{T}\tau$ symmetry operation, are provided in Supplementary Note S3, confirming that this staggered $\mathcal{T}$-odd response is the origin of the N\'eel SOT.
%--------- The "Punchline": Electrical Switching Criterion ---------

To quantify the switching potential, we calculate the net torque exerted on individual magnetic Fe1 and Fe2 sites. The $\mathcal{T}$-odd SEE response generates a staggered spin density that exerts an anti-damping-like N\'eel torque via the s-d exchange coupling ($J_{\text{sd}}$). This torque is quantified by an effective magnetic field, $\mathbf{B}_{T} \approx -(J_{\text{sd}}/\mu_B)\delta\mathbf{s}/M_s$, where $M_s \approx 2.3~\mu_B$ is the calculated local atomic magnetic moment. At the Fermi level, our calculations yield a substantial $\mathcal{T}$-odd susceptibility of $\chi_{yz} \approx 1.8 \times 10^{-10} \, \hbar\cdot\text{m/V}$. Under a moderate applied electric field of $10^6$~V/m, this generates a staggered spin density $|\delta\mathbf{s}|$ of approximately $1.6 \times 10^{-3} \, \mu_B/\text{nm}^3$. Assuming a typical exchange coupling of $J_{\text{sd}} \sim 1$~eV and a Gilbert damping of $\alpha_G \sim 0.01$~\cite{zeleznyRelativisticNeelOrderFields2014,PhysRevLett.132.136701}, the effective switching field $B_T/\alpha_G$ is estimated to be $\sim 270$~T. This corresponds to an effective energy scale of $\sim 36$~meV, which decisively exceeds the calculated out-of-plane magnetic anisotropy energy of 3.4~meV. Similar effective fields are obtained for the remaining Fe sites, demonstrating that the anti-damping torque generated in Fe-WS$_2$ is robust and sufficient to overcome the anisotropic energy barrier, enabling efficient electrical switching of the N\'eel vector.

\subsection{Machine-learning-accelerated discovery in partially intercalated systems}\label{sec:ML}

Having validated our symmetry-guided discovery framework on fully intercalated systems, we now extend this paradigm to more complex and experimentally synthesizable regime of partial intercalation\cite{zhou2025cation}. The arrangement of intercalants acts as a controllable source for symmetry breaking, creating structural prototypes that are inaccessible in the fully intercalated limit. However, the expanded configuration space leads to a combinatorial explosion of possible intercalant arrangements. To create a tractable subset for the high-throughput study, we restrict our enumeration to high-symmetry prototypes only, as detailed in the Supplementary Note S1.3. Even with this constraint, we identify 76 inequivalent parent structures for 1/2 concentration and 45 for 1/4 concentration. When combined with the extensive compositional space, this results in approximately $\sim$102,000 distinct configurations. This vast high-symmetry landscape renders a systematic DFT-based search computationally intractable.

\begin{figure}[htb]
	\centering
	\includegraphics[width=\linewidth]{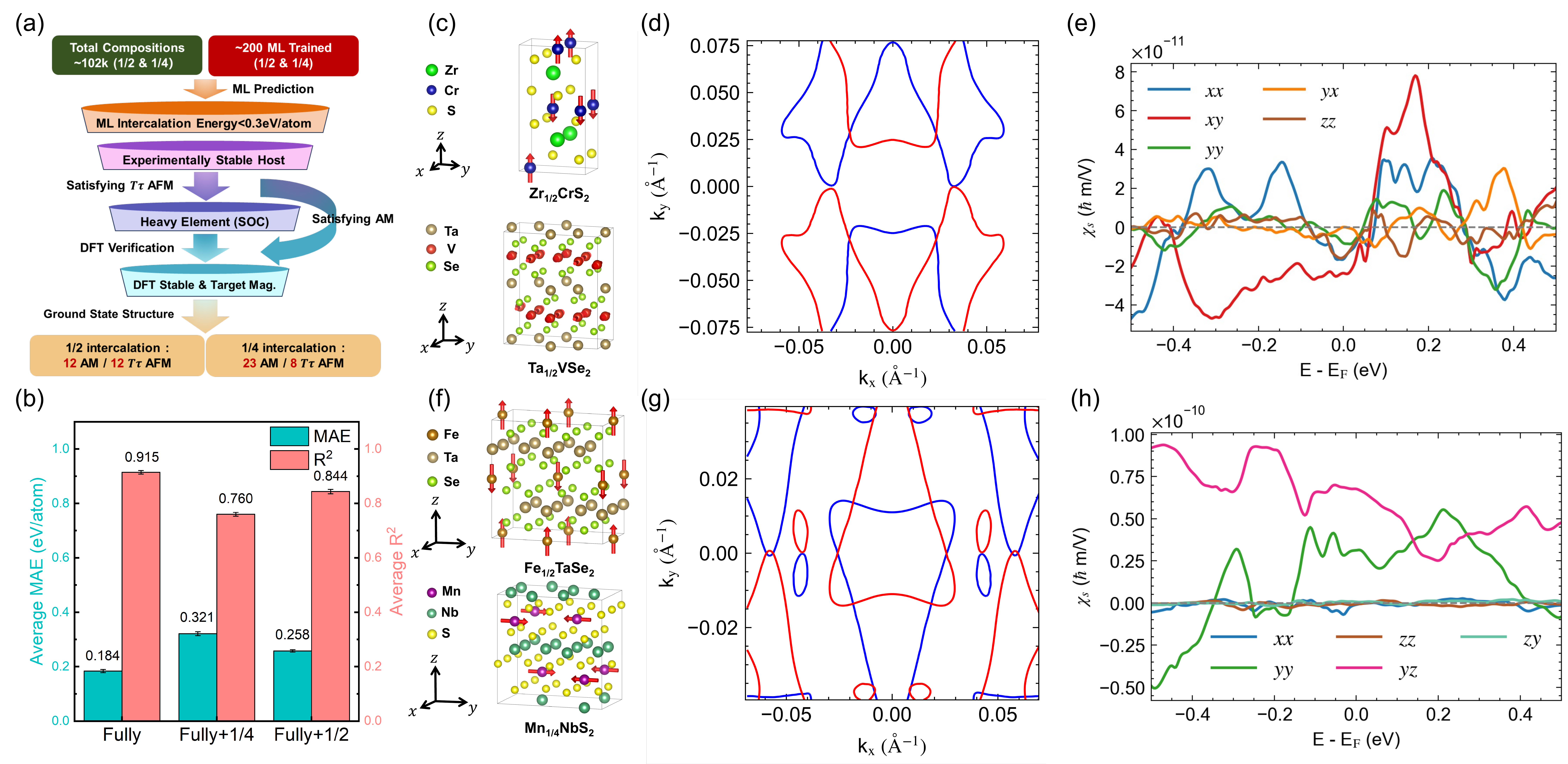}
    \caption{\textbf{Machine-learning-accelerated discovery of target magnetic phases in partially iTMDs.} 
    \textbf{a}, The hierarchical screening for stable partially iTMDs AM or $\mathcal{T}\tau$-AFM from a vast configuration space of \textasciitilde102,000 compositions.
    \textbf{b}, Statistical performance of our ML model. The bar chart displays the mean absolute error (MAE, cyan) and coefficient of determination ($R^2$, pink) for the model.
    \textbf{c, f}, Predicted stable crystal structures of representative AMs (top) and $\mathcal{T}\tau$-AFMs (bottom) identified by our screen at (\textbf{c}) 1/2 and (\textbf{f}) 1/4 concentrations.
    \textbf{d, g}, Spin-resolved iso-energy surfaces for (\textbf{d}) $\mathrm{Zr_{1/2}CrS_2}$ and (\textbf{g}) $\mathrm{Fe_{1/4}TaSe_2}$, both exhibiting characteristic d-wave nodal planes.
    \textbf{e, h}, Calculated $\mathcal{T}$-odd components of the local spin Edelstein effect susceptibility for the identified $\mathcal{T}\tau$-AFMs, $\mathrm{Ta_{1/2}VSe_2}$ (\textbf{e}) and $\mathrm{Mn_{1/4}NbS_2}$ (\textbf{h}), demonstrating significant response components conducive to electrical switching.}
	\label{fig:4}
\end{figure}

To address this challenge, we implement a data-efficient discovery strategy powered by our in-house developed MT-GNN based on DimeNet++ and mixture density networks~\cite{liang2024multi}. MT-GNN is designed to simultaneously infer intercalation energy and relax atomic structures, enabling robust generalization across distinct chemical environments. After being pre-trained on the dataset of fully intercalated systems, the model is transferred to partially iTMDs with either 1/2 or 1/4 concentration by including randomly chosen 100 structures from each case to the dataset for training. The performance of this model is illustrated in Fig.~\ref{fig:4}b, showing low mean absolute errors (MAE) of 0.258 eV/atom (1/2 concentration) and 0.321 eV/atom (1/4 concentration) on unseen test sets. 

With this validated machine learning (ML) framework, we proceed to execute the hierarchical screening workflow displayed in Fig.~\ref{fig:4}a. First, we screen for structures with ML-predicted intercalation energies below 0.3 eV/atom and consider only host materials that are experimentally synthesizable (Supplementary Note S4). 
Within this subset, we identify structures satisfying the specific symmetry requirements for AM or $\mathcal{T}\tau$-AFM orders. For the latter, we focus on systems containing heavy elements to ensure the strong SOC required for efficient switching. This multi-step screening process reduces the initial $\sim10^5$ candidates to $\sim2,000$ promising structures. Finally, restricting our analysis to candidates with DFT intercalation energies $< 0.3$~eV/atom, we perform DFT+U calculations to verify the AM or $\mathcal{T}\tau$-AFM ground-state order. This process yields 12 AM and 12 $\mathcal{T}\tau$-AFM candidates at 1/2 concentration, and 23 AM and 8 $\mathcal{T}\tau$-AFM candidates at 1/4 concentration (Table~\ref{table2}). Additionally, our screening reveals a broader set of 119 AM and 129 $\mathcal{T}\tau$-AFM metastable candidates meeting the thermodynamic stability threshold, which are listed in Supplementary Table S10-11.

As an illustration for the 1/2 concentration case, we predict a d-wave altermagnetic phase in Zr-intercalated CrS$_2$ (Fig.~\ref{fig:4}c). Its iso-energy surface near the Fermi level (Fig.~\ref{fig:4}d) clearly exhibits a two-fold rotational symmetry, a definitive breaking of the underlying C$_3$ symmetry and a hallmark of a d-wave state. Interestingly, the mechanism here is distinct from our previous example, as the C$_3$ symmetry is broken by the arrangement order of the Zr intercalants, rather than by the magnetic order, showcasing a new design route to d-wave altermagnets. For $\mathcal{T}\tau$-AFM, we identify Ta-intercalated VSe$_2$ as a promising candidate, which possesses a $m.1'$ magnetic point group. The calculated local $\mathcal{T}$-odd SEE susceptibility projected onto a magnetic V atom (Fig.~\ref{fig:4}e) reveals a giant response near the Fermi level, dominated by the $\chi_{xx}$ and $\chi_{xy}$ components. The material exhibits a remarkably small in-plane magnetic anisotropy energy ($\sim 0.4$~meV) compared to its out-of-plane magnetic anisotropy energy ($\sim 1.3$~meV). This low barrier, combined with the giant SEE response, indicates its potential for efficient electrical switching, as confirmed by quantitative analysis in the Supplementary Note S5.

The success of our framework is best exemplified at 1/4 concentration by the identification of two recently synthesized altermagnets, $\mathrm{Co_{1/4}TaSe_2}$ and $\mathrm{Co_{1/4}NbSe_2}$~\cite{spragueObservationAltermagneticSpin2025,daleNonrelativisticSpinSplitting2024,grahamLocalProbeEvidence2025,sakhyaElectronicStructureLayered2025,regmiAltermagnetismLayeredIntercalated2025}. Our predicted structures and g-wave Fermi surfaces for these compounds are in excellent agreement with the literature, as demonstrated in the Supplementary Note S6. In addition, we highlight a predicted d-wave candidate shown in the upper panel of Fig.~\ref{fig:4}f, $\mathrm{Fe_{1/4}TaSe_2}$. Although derived from the same fully intercalated structural prototype (H-AA'-octa) as the experimental Co-containing compounds, the Fe intercalants occupy distinct octahedral sites that break the $C_3$ rotational symmetry, resulting in a d-wave state (Fig.~\ref{fig:4}g). As for $\mathcal{T}\tau$-AFM materials, we showcase a candidate, $\mathrm{Mn_{1/4}NbS_2}$, possessing the $mm2.1'$ magnetic point group. The calculated local $\mathcal{T}$-odd SEE susceptibility for the Mn site reveals a particularly large response in the $\chi_{yy}$ and $\chi_{yz}$ component (Fig.~\ref{fig:4}h), consistent with the non-zero tensor elements allowed by symmetry. Comprehensive calculations detailed in the Supplementary Note S5, including full site-resolved analysis and effective switching field estimations, confirm its potential for efficient electrical switching. 

\section{Discussion}\label{sec3}
In this work, we establish a symmetry-guided and AI-accelerated computational framework to systematically navigate the complex landscape of iTMDs for AFM spintronics applications. The significance of our AI-accelerated paradigm lies not only in its efficiency but also in its ability to connect with and advance beyond current experimental investigation. The prediction of the experimentally synthesized g-wave altermagnets $\mathrm{Co_{1/4}TaSe_2}$ and $\mathrm{Co_{1/4}NbSe_2}$ serves as the validation of our approach's reliability. More importantly, our search reveals a far richer structural and magnetic landscape, leading to the discovery of 219 altermagnets in diverse configurations. These include many d-wave altermagnets, such as $\mathrm{Zr_{1/2}CrS_2}$ and $\mathrm{Fe_{1/4}TaSe_2}$, whose thermodynamic stability suggests the feasibility for experimental synthesis. These d-wave altermagnets introduce the technologically advantageous spin-splitter effect, that is absent in g-wave systems. Meanwhile, we identify 243 $\mathcal{T}\tau$-AFM candidates where $\mathcal{T}$-odd Néel SOTs can exist in both metallic and insulating systems. This overcomes the limitations of $\mathcal{P}\mathcal{T}$-symmetric systems, where $\mathcal{T}$-even torques are typically confined to metallic systems, thereby unlocking a material repository for versatile low-power SOT switching applications.
% anti-damping torques that are intrinsically more efficient than the field-like torques in conventional $P\mathcal{T}$-symmetric AFMs, indicating a clear path toward ultralow-power spintronic devices.

While our current study focuses on bulk hosts, the symmetry-guided framework established here can be easily generalized to different layered material families, such as MXenes, square-lattice compounds, and 2D intercalation compounds\cite{zhaoEngineeringCovalentlyBonded2020}. Furthermore, the search can be generalized to diverse quantum materials with different symmetries, such as multiferroics, nonlinear transport phenomena\cite{zhaoGeneralTheoryLongitudinal2024,kaplanUnificationNonlinearAnomalous2024}, and even systems hosting non-collinear magnetic textures. This represents a particularly exciting opportunity, as recent work has shown that non-collinear magnets can generate non-relativistic torques arising from the magnetic texture itself, providing an alternative pathway to electrical control that does not rely on SOC~\cite{gonzalez-hernandezNonrelativisticTorqueEdelstein2024}. We believe our findings will facilitate the experimental exploration of the identified candidates, paving the way for the development of next-generation spintronic devices.

%TC:ignore
\section{Materials and Methods}\label{sec11}
\subsection*{First-principles Calculations}
All first-principles calculations are performed using the Vienna Ab initio Simulation Package \cite{kresseEfficientIterativeSchemes1996} based on DFT. The projector augmented-wave method is employed to describe the interaction between core and valence electrons\cite{kresseUltrasoftPseudopotentialsProjector1999}. Structural relaxations for all candidate materials are carried out using the Perdew-Burke-Ernzerhof (PBE) exchange-correlation functional\cite{perdewGeneralizedGradientApproximation1996} until the forces on all atoms are below 0.02~eV/\AA{}. For the subsequent self-consistent electronic structure and magnetic property calculations, the PBE+U method\cite{PhysRevB.57.1505} is employed to describe electron correlation in the d-orbitals of the transition metal atoms. The specific Hubbard U and exchange J parameters used for each element are detailed in the Supplementary Note S7. A plane-wave basis set with a default energy cutoff is used. The k-point meshes for Brillouin zone integration are generated using the Monkhorst-Pack scheme\cite{monkhorst1976special}, with the smallest allowed spacing between k-points set to 0.3~\AA$^{-1}$.
\subsection*{Symmetry Analysis}

The analysis of magnetic symmetry is performed using a combination of specialized software packages. The magnetic space groups for the crystal structures are determined using the \texttt{findsym} code\cite{stokes2005findsym}. The identification of altermagnetic candidates is carried out using the \texttt{amcheck} software package\cite{10.21468/SciPostPhysCodeb.30}. For a detailed analysis of the spin group operations, we employ the \texttt{findspingroup} code\cite{chenEnumerationRepresentationTheory2024}.

\subsection*{Machine Learning Model for Energy Prediction}

To efficiently explore the large structural space of partially intercalated systems, we employ a multi-task Graph Neural Network~\cite{liang2024multi} designed to simultaneously predict intercalation energy and relaxed atomic structures. We utilize Latin Hypercube Sampling to generate a candidate pool of $\sim500$ structures per concentration, uniformly sampling stacking order, host composition, and intercalant species. By excluding structures with significant distortion after DFT calculations, 279 and 352 valid structures are obtained for the 1/2 and 1/4 concentrations, respectively. From these small datasets, we randomly choose 100 structures for each concentration to be included in the full-intercalation dataset (No. of data $N=5,344$), forming composite sets for training. The remaining valid structures serve as the test set to evaluate model generalization. By training on these composite datasets, the models achieve high accuracy with MAE $\approx 0.26$--$0.32$ eV/atom. The trained model is used to predict intercalation energies for the configurations of 1/2 and 1/4 concentrations, which comprise 63,811 and 38,164 distinct structures, respectively. A complete description of the model architecture and dataset splitting is provided in the Supplementary Note S8 and Tables S3-S4.

\subsection*{Transport Property Calculations}

\textbf{Spin Conductivity.} The spin conductivity tensor ($\sigma^s$) responsible for the spin-splitter effect is calculated using the Boltzmann transport equation under the constant relaxation time approximation, as implemented in the \texttt{Wannier90} package\cite{pizzi2020wannier90,pizzi2014boltzwann}. 
\begin{equation}
\sigma_{i j}^{s}\left(\epsilon_{F}\right)=-\frac{e^{2} \tau}{8 \pi^{3} \hbar^{2}} \sum_{n} \int \frac{\partial \epsilon_{n \mathbf{k}}^{s}}{\partial k_{i}} \frac{\partial \epsilon_{n \mathbf{k}}^{s}}{\partial k_{j}} \frac{\partial f^{0}}{\partial \epsilon_{n \mathbf{k}}^{s}} d^{3} k
\end{equation}
where $\tau$ is the constant relaxation time, $\epsilon_{n \mathbf{k}}^{s}$ is the energy eigenvalue for an electron with spin $s$ in band $n$, and $f^{0}$ is the Fermi-Dirac distribution function. The resulting conductivity tensor, $\sigma_{ij}^s$, is symmetric. 
To ensure convergence, a dense k-point mesh of 250$\times$400$\times$120 is used for the Brillouin zone integration.

\textbf{Spin Edelstein Effect.} The linear response tensor for the spin Edelstein effect ($\chi_{ij}$) is calculated using the Kubo formula as implemented in the \texttt{Linres} code\cite{ZeleznyjWannierlinearresponseBitbucket}. The susceptibility tensor is decomposed into $\mathcal{T}$-even and $\mathcal{T}$-odd components\cite{freimuthSpinorbitTorquesCo2014,liIntrabandInterbandSpinorbit2015}:
\begin{equation}
\chi_{i j}^{\text{even}} = -\frac{e \hbar}{\pi} \sum_{\mathbf{k}, m, n} \frac{\operatorname{Re}\left[\left\langle\Psi_{\mathbf{k} n}\right| \hat{S}_{i}\left|\Psi_{\mathbf{k} m}\right\rangle\left\langle\Psi_{\mathbf{k} m}\right| \hat{v}_{j}\left|\Psi_{\mathbf{k} n}\right\rangle\right] \Gamma^{2}}{\left(\left(\varepsilon_{F}-\varepsilon_{\mathbf{k} n}\right)^{2}+\Gamma^{2}\right)\left(\left(\varepsilon_{F}-\varepsilon_{\mathbf{k} m}\right)^{2}+\Gamma^{2}\right)}
\end{equation}
\begin{equation}
\chi_{i j}^{\text{odd}} = 2 e \hbar \sum_{\mathbf{k}, n \neq m} \operatorname{Im}\left[\left\langle\psi_{n \mathbf{k}}\right| \hat{S}_{i}\left|\psi_{m \mathbf{k}}\right\rangle\left\langle\psi_{m \mathbf{k}}\right| \hat{v}_{j}\left|\psi_{n \mathbf{k}}\right\rangle\right]
\times \frac{\Gamma^{2}-\left(\varepsilon_{\mathbf{k} n}-\varepsilon_{\mathbf{k} m}\right)^{2}}{\left[\left(\varepsilon_{\mathbf{k} n}-\varepsilon_{\mathbf{k} m}\right)^{2}+\Gamma^{2}\right]^{2}}
\end{equation}
Here, $\psi_{n\mathbf{k}}$ and $\varepsilon_{n\mathbf{k}}$ are the Bloch wavefunctions and eigenvalues, respectively, $\hat{S}_i$ is the total spin operator, $\hat{v}_j$ is the velocity operator, and $\Gamma$ is a phenomenological broadening parameter related to the relaxation time ($\tau = \hbar/2\Gamma$), which is set to 0.01~eV in our calculations. These $\mathcal{T}$-even and $\mathcal{T}$-odd components are associated with intraband and interband contributions, respectively. To compute the local Edelstein effect arising from a specific magnetic sublattice or atomic site, as presented in our main text, the total spin operator $\hat{S}_i$ in the formulae above is replaced by a corresponding projection operator for the considered sublattice or site.% For the calculation of the SEE tensor for the representative candidate in Fig.~\ref{SOTdemo}, a k-point mesh of 400$\times$230$\times$100 is used.
%TC:endignore

%%=============================================================%%
%%  后记 (Back Matter)
%%=============================================================%%
%TC:ignore
% \backmatter

% % \bmhead{Supplementary information}
% Supplementary information is available for this paper at https://doi.org/10.1038/...

\bibliography{ref} % 您的 .bib 文件名 (不带后缀)
\bibliographystyle{sciencemag}

\section*{Acknowledgments}
\paragraph*{Funding:}
This work was supported by the National Natural Science Foundation of China (Project No. 12474236), and the Shenzhen Science and Technology Program (ZDSYS20230626091100001).

% \section*{Declarations}
% \begin{itemize}
% \item Funding: the National Natural Science Foundation of China (Project No. 12474236), and the Shenzhen Science and Technology Program (ZDSYS20230626091100001)
\paragraph*{Competing interests:}
There are no competing interests to declare.
\paragraph*{Data and materials availability:}
The data that support the findings of this study are available from the corresponding author upon reasonable request. The code for the machine learning portion of this work is available at https://github.com/groupzou.
\paragraph*{Author contributions:}
X.L.Z. conceived and supervised the project. Y.P. performed the symmetry analysis, constructed and calculated the magnetic structures, and calculated the electronic and transport properties. R.S.Z. trained and applied the machine learning models. L.Y.Z. performed all structural relaxations. Y.P., Y.G., and L.Y.Z. contributed to the construction of the intercalated structures. X.L.Z., Y.P., Y.G., R.S.Z., and L.Y.Z. analyzed the results and discussed with X.B.C. and W.H.D. Y.P. wrote the initial draft of the manuscript, with contributions from Y.G. and R.S.Z. All authors contributed to the discussions of the results and the preparation and revision of the final manuscript. Y.P., Y.G., R.S.Z., and L.Y.Z contributed equally to this work. 
% \end{itemize}
\subsection*{Supplementary materials}
Materials and Methods\\
Supplementary Text\\
Figs. S1 to S15\\
Tables S1 to S11\\
% References \textit{\cite{spragueObservationAltermagneticSpin2025,regmiAltermagnetismLayeredIntercalated2025,liang2024multi,gasteiger_dimenetpp_2020}}
%TC:endignore

\newpage

\newpage

\newpage

\newpage

\clearpage
\begin{table*}[tb]
  \centering
  \caption{The 29 structural prototypes, comprising pristine hosts and fully intercalated phases. Following symmetry analysis, all listed prototypes are candidates for hosting $\mathcal{T}\tau$ antiferromagnetism, while those potentially hosting altermagnetism are marked with an asterisk (*).}
  \label{tab:full_intercalation_prototypes}
  \begin{tabular}{l l l l l}
    \toprule
    \textbf{Phase} & \multicolumn{4}{c}{\textbf{Stacking and Intercalation Site}} \\
    \midrule
    \textbf{T-phase} & T-AA & T-AB & T-AC & T-AA'* \\
    & T-AB'* & T-AA'-prism* & T-AB'-tetra\_1* & T-AB'-tetra\_2 \\
    &T-AB'-octa* & T-AA-octa & T-AA-tetra & T-AB-octa  \\
    &T-AB-tetra  & T-AC-prism& & \\
    \textbf{H-phase} & H-AA & H-AB & H-AA' & H-AB' \\
    & H-AC' & H-AA'-octa* & H-AB'-octa* & H-AA'-tetra\_1* \\
    & H-AA'-tetra\_2* &H-AB'-tetra\_1* & H-AB'-tetra\_2* & H-AC'-prism* \\ 
    &  H-AA-prism & H-AB-octa &H-AB-tetra  &  \\
    \bottomrule
  \end{tabular}
\end{table*}

\begin{table*}[htb]
\centering
\caption{Summary of screened magnetic candidates at fractional concentrations. The table lists chemical formulas for stable altermagnetic and $\mathcal{T}\tau$-antiferromagnetic structures categorized by host phase and stacking types. Bold formulas indicate the lowest-energy configuration among all stacking types, while the symbol $^{\dagger}$ denotes candidates with positive intercalation energies below 0.3~eV/atom. Details are provided in Supplementary Note S9.}
\label{table2}
\label{tab:candidates_grid}

{\small 
\begin{tabular}{@{} p{0.06\linewidth} >{\raggedright\arraybackslash}p{0.44\linewidth} >{\raggedright\arraybackslash}p{0.44\linewidth} @{}}
\toprule
\textbf{Conc.} & \textbf{AM} & \textbf{$\mathcal{T}\tau$ AFM} \\
\midrule
\textbf{1/2} & \textbf{T-AA'/AB':}\newline
\makebox[0.495\linewidth][l]{$\mathrm{Cr_{1/2}CrS_{2}}$}\allowbreak
\makebox[0.495\linewidth][l]{$\mathbf{Fe_{1/2}NbS_{2}}$}\allowbreak
\makebox[0.495\linewidth][l]{$\mathbf{Hf_{1/2}VSe_{2}}$}\allowbreak
\makebox[0.495\linewidth][l]{$\mathbf{Mn_{1/2}CrS_{2}}$}\allowbreak
\makebox[0.495\linewidth][l]{$\mathbf{Mn_{1/2}NbS_{2}}$}\allowbreak
\makebox[0.495\linewidth][l]{$\mathrm{Nb_{1/2}CrS_{2}}$}\allowbreak
\makebox[0.495\linewidth][l]{$\mathbf{Ta_{1/2}CrS_{2}}$}\allowbreak
\makebox[0.495\linewidth][l]{$\mathrm{Ti_{1/2}FeSe_{2}}$}\allowbreak
\makebox[0.495\linewidth][l]{$\mathbf{Zr_{1/2}CrS_{2}}$}\allowbreak
\newline

\textbf{H-AA'/AB'/AC':}\newline
\makebox[0.495\linewidth][l]{$\mathbf{Cu_{1/2}VS_{2}}$}\allowbreak
\makebox[0.495\linewidth][l]{$\mathbf{Ni_{1/2}NbS_{2}}$}\allowbreak
\makebox[0.495\linewidth][l]{$\mathbf{Ni_{1/2}NbTe_{2}}$}\allowbreak & \textbf{T-AA/AB/AC:}\newline
\makebox[0.495\linewidth][l]{$\mathrm{Co_{1/2}MoS_{2}}$}\allowbreak
\makebox[0.495\linewidth][l]{$\mathbf{Co_{1/2}NbS_{2}}$}\allowbreak
\makebox[0.495\linewidth][l]{$\mathrm{Fe_{1/2}NbS_{2}}$}\allowbreak
\makebox[0.495\linewidth][l]{$\mathrm{Hf_{1/2}FeSe_{2}}$}\allowbreak
\makebox[0.495\linewidth][l]{$\mathrm{Mn_{1/2}NbS_{2}}$}\allowbreak
\makebox[0.495\linewidth][l]{$\mathbf{Mn_{1/2}TaS_{2}^{\dagger}}$}\allowbreak
\newline

\textbf{H-AA/AB:}\newline
\makebox[0.495\linewidth][l]{$\mathrm{Mn_{1/2}NbS_{2}}$}\allowbreak
\makebox[0.495\linewidth][l]{$\mathrm{Ta_{1/2}VS_{2}^{\dagger}}$}\allowbreak
\newline

\textbf{H-AA'/AB'/AC':}\newline
\makebox[0.495\linewidth][l]{$\mathbf{Cu_{1/2}VSe_{2}^{\dagger}}$}\allowbreak
\makebox[0.495\linewidth][l]{$\mathbf{Nb_{1/2}VSe_{2}}$}\allowbreak
\makebox[0.495\linewidth][l]{$\mathbf{Pt_{1/2}VS_{2}^{\dagger}}$}\allowbreak
\makebox[0.495\linewidth][l]{$\mathbf{Ta_{1/2}VSe_{2}}$}\allowbreak \\
\midrule
\textbf{1/4} & \textbf{T-AA'/AB':}\newline
\makebox[0.495\linewidth][l]{$\mathbf{Cr_{1/4}VS_{2}}$}\allowbreak
\makebox[0.495\linewidth][l]{$\mathbf{Fe_{1/4}VS_{2}}$}\allowbreak
\makebox[0.495\linewidth][l]{$\mathbf{Hf_{1/4}FeTe_{2}}$}\allowbreak
\makebox[0.495\linewidth][l]{$\mathbf{Mo_{1/4}CrS_{2}}$}\allowbreak
\makebox[0.495\linewidth][l]{$\mathbf{Nb_{1/4}CrS_{2}}$}\allowbreak
\makebox[0.495\linewidth][l]{$\mathbf{Ta_{1/4}CrS_{2}}$}\allowbreak
\makebox[0.495\linewidth][l]{$\mathbf{V_{1/4}VS_{2}}$}\allowbreak
\makebox[0.495\linewidth][l]{$\mathbf{W_{1/4}CrS_{2}}$}\allowbreak
\newline

\textbf{H-AA'/AB'/AC':}\newline
\makebox[0.495\linewidth][l]{$\mathbf{Co_{1/4}NbS_{2}}$}\allowbreak
\makebox[0.495\linewidth][l]{$\mathbf{Co_{1/4}NbSe_{2}}$}\allowbreak
\makebox[0.495\linewidth][l]{$\mathbf{Co_{1/4}TaS_{2}}$}\allowbreak
\makebox[0.495\linewidth][l]{$\mathbf{Co_{1/4}TaSe_{2}}$}\allowbreak
\makebox[0.495\linewidth][l]{$\mathbf{Co_{1/4}VS_{2}}$}\allowbreak
\makebox[0.495\linewidth][l]{$\mathbf{Fe_{1/4}NbS_{2}}$}\allowbreak
\makebox[0.495\linewidth][l]{$\mathbf{Fe_{1/4}NbSe_{2}}$}\allowbreak
\makebox[0.495\linewidth][l]{$\mathbf{Fe_{1/4}TaSe_{2}}$}\allowbreak
\makebox[0.495\linewidth][l]{$\mathbf{Fe_{1/4}VS_{2}}$}\allowbreak
\makebox[0.495\linewidth][l]{$\mathbf{Mn_{1/4}NbSe_{2}}$}\allowbreak
\makebox[0.495\linewidth][l]{$\mathbf{Mo_{1/4}VS_{2}}$}\allowbreak
\makebox[0.495\linewidth][l]{$\mathbf{Ni_{1/4}VSe_{2}}$}\allowbreak
\makebox[0.495\linewidth][l]{$\mathbf{V_{1/4}NbS_{2}}$}\allowbreak
\makebox[0.495\linewidth][l]{$\mathbf{V_{1/4}NbSe_{2}}$}\allowbreak
\makebox[0.495\linewidth][l]{$\mathbf{Zn_{1/4}VS_{2}}$}\allowbreak & \textbf{T-AA/AB/AC:}\newline
\makebox[0.495\linewidth][l]{$\mathrm{Cr_{1/4}NbS_{2}}$}\allowbreak
\makebox[0.495\linewidth][l]{$\mathbf{Pd_{1/4}VS_{2}}$}\allowbreak
\makebox[0.495\linewidth][l]{$\mathrm{V_{1/4}NbS_{2}}$}\allowbreak
\newline

\textbf{T-AA'/AB':}\newline
\makebox[0.495\linewidth][l]{$\mathbf{Hf_{1/4}VS_{2}}$}\allowbreak
\makebox[0.495\linewidth][l]{$\mathrm{Mn_{1/4}NbS_{2}}$}\allowbreak
\makebox[0.495\linewidth][l]{$\mathrm{Pd_{1/4}VS_{2}^{\dagger}}$}\allowbreak
\makebox[0.495\linewidth][l]{$\mathbf{Ti_{1/4}VSe_{2}}$}\allowbreak
\makebox[0.495\linewidth][l]{$\mathbf{Zr_{1/4}VS_{2}}$}\allowbreak \\
\bottomrule
\end{tabular}
}
\end{table*}

\end{document}